\documentclass[12pt,preprint]{aastex}

\usepackage{url}

\begin{document}

\title{Extending Counter-Streaming Motion from an Active Region Filament to Sunspot Light Bridge}

\author{Haimin Wang$^{1,3}$, Rui Liu$^2$, Qin Li$^{1,3}$, Chang Liu$^{1,3}$,  Na Deng$^{1,3}$,  Yan Xu$^{1,3}$,  Ju Jing$^{1,3}$, Yuming Wang$^{2}$ and Wenda Cao$^3$}

\affil{1. Space Weather Research Laboratory, New Jersey Institute of Technology, University Heights, Newark, NJ 07102-1982, USA}

\affil{2. CAS Key Laboratory of Geospace Environment, Department of Geophysics and Planetary Sciences, University of Science and Technology of China, Hefei 230026, China}

\affil{3. Big Bear Solar Observatory, New Jersey Institute of Technology, 40386 North Shore Lane, Big Bear City, CA 92314-9672, USA}

\email{haimin.wang@njit.edu}

\begin{abstract}

We analyze the high-resolution observations from the 1.6~m  telescope at Big Bear Solar Observatory that cover an active region filament.  Counter-streaming motions are clearly observed in the filament. The northern end of the counter-streaming motions extends to a light bridge, forming a spectacular circulation pattern around a sunspot, with clockwise motion in the blue wing and counterclockwise motion in the red wing as observed in H$\alpha$ off-bands. The apparent speed of the flow is around 10 to 60 km~s$^{-1}$ in the filament, decreasing to 5 to 20 km~s$^{-1}$ in the light bridge.  The most intriguing results are the magnetic structure and the counter-streaming motions in the light bridge. Similar to those in the filament, magnetic fields show a dominant transverse component in the light bridge. However, the filament is located between opposite magnetic polarities, while the light bridge is between strong fields of the same polarity. We analyze the power of oscillations with the image sequences of constructed Dopplergrams, and find that the filament's counter-streaming motion is due to physical mass motion along fibrils,  while the light bridge's counter-streaming motion is due to oscillation in the direction along the line-of-sight. The oscillation power peaks around 4 minutes.  However,  the section of the light bridge next to the filament also contains a component of the extension of the filament in combination with the oscillation, indicating that some strands of the filament are extended to and rooted in that part of the light bridge.

\end{abstract}

\keywords{Sun: activity  -- Sun: filaments -- Sun: magnetic fields }

\section{INTRODUCTION}

Solar filaments are cool and dense material suspended in the solar chromosphere and  corona.  Beyond the limb,   filaments appear as bright prominences against a dark background (Tandberg-Hanssen, 1995).   Filaments' structure and evolution are of significance in the study of solar activities, as their eruptions are closely associated with solar flares and coronal mass ejections (CMEs)  (Shibata \& Magara, 2011;  Webb \& Howard, 2012).   In particular, the two-ribbon structure of flares is often observed as filaments rise through arcade fields that subsequently reconnect. Two ribbons are associated with opposite magnetic polarities, and both run parallel to the magnetic polarity inversion line (PIL) lying between them. Such a configuration can be explained by the classical reconnection model called the CSHKP model (Carmichael 1964; Sturrock 1966; Hirayama 1974; Kopp \& Pneuman 1976).

Besides the solar eruptions,  filaments contain rich dynamic structure and plasma flow motions along the filament threads.  Among the most significant ones is the  counter-streaming motion (Zirker et al., 1998),  where the opposite directions of flows co-exist in filaments. Earlier investigations indicated that the counter-streaming is due to longitudinal oscillations of thin threads, restored by gravitational force (e.g., Lin et al., 2003, Karpen et al., 2006, Xia et al., 2011, Chen et al., 2014, Shen et al., 2015). Related to this,  a magnetic dip is required to support the filament and balance the gravity force of the filament.  We note that most of prior studies targeted at the quiescent filaments which are much higher than the active regions filaments. The state-of-the-art observations with high spatiotemporal resolution of the 1.6~m Goode Solar Telescope (GST,  formerly NST; Goode et al. 2010; Cao et al. 2010) at Big Bear Solar Observatory (BBSO) allows an assessment of the low atmospheric structure in an unprecedented detail.  As an example, Zou et al. (2016) studied GST data for an active region filament, and concluded that the filament is supported by sheared arcades and the counter-streaming motion is due to unidirectional flows with alternative directions.

On another seemingly unrelated topic, sunspot light bridges (LBs) also received significant attention in recent years with high resolution.  LBs typically have weaker yet highly inclined (Lites et al., 1991, Leka, 1997) magnetic fields comparing with surrounding umbrae.  Those fields are formed by large-scale convective flows (Torium et al., 2015a,b, Felipe et al., 2016). In this regard, the fields in the LBs bear some resemblance to those of filaments, i.e., horizontal fields surrounded by the nearby more vertical fields. The difference is also very obvious:  filaments are often aligned with PILs, between two sides of the opposite magnetic fields;  the magnetic fields in the two sides of LBs usually have the same magnetic polarity, except those LBs dividing $\delta$ sunspots.  Recently, the sunspot light walls are discussed to be associated with oscillation motions above LBs (e.g. Yang et al., 2015, Zhang et al., 2017,  Hou et al., 2017).  It is interesting to find if chromospheric materials above LBs have similar counter-streaming effects as observed in filaments.

In this Letter, we report the high-resolution observations of  NOAA active region (AR) 12371, on 2015 June 20, in which a filament is covered by comprehensive observations. Several sections of LBs are visible in this AR, including a section seemingly connecting to the filament.  We focus on the study of magnetic and flows in the filament and the connected LB.  The study uses the multi-wavelength observations from GST and the Solar Dynamic Observatory (SDO, Pesnell et al., 2012),  with the aid of the nonlinear force-free field (NLFFF) extrapolation.

\section{OBSERVATIONS AND DATA PROCESSING}

With a 308-element adaptive optics system and speckle-masking image reconstruction using 100 frames, BBSO/GST achieved diffraction-limited imaging in TiO band (a proxy for photospheric continuum at 7057~\AA) and in H$\alpha$ at five wavelengths: $-$1.0~\AA, $-$0.6~\AA, line center, +0.6~\AA, and +1.0~\AA\ on 2015 June 20.  The target was the core area of NOAA AR 12371.  The  pixel scale of above data sets is  0.\arcsec03, and the cadence is nominally 30 s.  The GST observations covered about six hours under excellent and consistent seeing condition.

The full-disk vector magnetic field data with a 12 minute cadence from the Helioseismic and Magnetic Imager (HMI; Schou et al. 2012) on board SDO are derived using the VFISV inversion code by Borrero et al. (2011). The accuracy of HMI magnetograms is 10 G for line-of-sight fields and 100 G for transverse fields. After preprocessing the photospheric boundary to best suit the force-free condition (Wiegelmann et al. 2006), we constructed NLFFF models using the ``weighted optimization'' method (Wiegelmann 2004) with the error treatment incorporated (Wiegelmann \& Inhester 2010; Wiegelmann et al. 2012). Ancillary data in 1700~\AA\ from the Atmospheric Imaging Assembly (AIA; Lemen et al. 2012) on board SDO are additionally used.

The images at different wavelengths are aligned with feature matching. TiO images are easily aligned with H$\alpha$ far wings and continuum images of HMI and AIA. HMI magnetograms are then aligned with reference of HMI continuum.   The alignment accuracy is less than 1\arcsec.

\section{RESULTS}

In Figure 1, we show the context images to define different areas under study.  The active region filament (marked with the letter F) is shown clearly in the H$\alpha$ line center image. Its two footpoints are marked as SF (Southern Footpoint) and NF (Northern Footpoint). Viewing larger scale structures from full-disk H$\alpha$ images, this filament is one leg of a sigmoid structure, with the SF lying towards the center of the sigmoid. In the corresponding TiO image,  we divide the region surrounding the top sunspot into three sections, Light Bridge section 1 (LB1),  Light Bridge section 2 (LB2),  and a penumbral region corresponding to the filament (F) in H$\alpha$.  In the bottom part of the figure we show H$\alpha$$-$0.6~\AA\ and +0.6~\AA\ images.   The flow motions under study are most obvious in these wavelengths.  The accompanied movies of H$\alpha$$-$0.6~\AA\ and +0.6~\AA\ demonstrate the circular counter-streaming motions  spanning from the filament to the LB (LB1 and LB2). The blue-wing movie shows clock-wise motion,  while the red wing, counter-clock-wise.

We use two methods to measure the speed of the flow motion. Figure 2 shows the results of local correlation tracking (LCT).  Although in general,  the motion direction matches with visual inspection of the movies, the derived magnitude of the counter-streaming motion in the order of 1 km~s$^{-1}$ is substantially underestimated. The reason is that the flow motions only exist along certain fibrils,  while the LCT gives an average flow speed.  Therefore,  we use time-slice method to manually measure the flow speed along certain prominent fibrils (see Li et al, 2017).  Figure  3 shows the results.  Based on the measurements of about 100 moving features, the flow speed is 20 to 60 km~s$^{-1}$ in the section of the filament and is reduced to 5 to 20 km~s$^{-1}$ in the LB.  The moving fibrils are substantially longer in the filament (as long as 30\arcsec,   compared to 1--3\arcsec\ in the LB).  The speed in the red wing is higher than that in the blue wing in the filament (40 vs. 20 km~s$^{-1}$) and LB1 (20 vs. 10 km~s$^{-1}$). In LB2, the mean speed is about 5 km~s$^{-1}$ for both wings.

To distinguish if the observed flow motion is a mass motion or a signature of chromospheric oscillations of filament threads,  we construct pseudo Dopplergrams using the difference between red and blue wing images of $\pm$0.6~\AA.  Wavelet analysis is then performed on the time sequence of Dopplergrams to derive the power of possible periodic motions. The wavelet power spectra  are calculated by using the Morlet wavelet analysis technique of Torrence \& Compo (1998), which provides excellent time and frequency localization information. In general,  throughout the AR, the oscillation power peaks at the period around 4 minutes (4.2 mHz),   which is between the photospheric p-mode and 3-minute chromospheric oscillation.
In Figure 4,  we plot the power of  oscillation along a curve connecting F, LB1 and LB2.   The mean 4-minute oscillating powers are 0.0014, 0.0063 and 0.021 in F, LB1, and LB2, respectively. It is obvious that in the filament,  the oscillation power is at least one order of magnitude smaller than that in LB2.   Therefore, we conclude that the mass flow is the mechanism for the counter-streaming motion in the filament, while the oscillation is the likely cause of the apparent counter-streaming motion in LB2. Notably, LB1 is an interesting section.  It shows short, hairy, and somewhat Y-shaped patterns; meanwhile, some strands of the filament also extend to this section with substantially longer fibrils (in the size of about 2,000 to 3,000 km, in comparison to typical background size of 700 km in both LB1 and LB2).  It has a dominant component of oscillation with a period of 4--6 minutes, therefore the underlying physical mechanism is similar to that in LB2. However, compared to LB2 the oscillation power in LB1 is reduced by a factor of 3. We conclude that LB1 has a hybrid nature, as it combines the characteristics of oscillation motion and the mass motion extended from the filament.

The next step is to analyze the magnetic topology around the sunspot including the filament and the LB. Figure 5 shows the vector magnetic field plot with the HMI data, which demonstrates a clockwise circulation pattern of transverse field vectors from LB2 to LB1 and to F. The lower right panel of Figure 5 shows the result of NLFFF extrapolation that resembles the filament well.  The southern end point is well defined and matches with the tracing in H$\alpha$.  The northern end of the filament does not match the H$\alpha$ tracing well.   We believe that the NLFFF extrapolation emphasizes magnetic connectivity at higher altitudes where the NLFFF conditions are better satisfied, while the H$\alpha$ observation shows the lower atmospheric structure.  Therefore,  the trajectory of continuous motion in the chromosphere is not reflected in the traced NLFFF flux rope from the filament to LB1. On the other hand, the H$\alpha$ material might trace only certain segments of field lines.

\section{SUMMARY AND DISCUSSION}

We have presented a detailed study of flows and magnetic fields around a sunspot in AR 12371.  Here are the key results:

\begin{enumerate}

\item This is the first time that counter-streaming motions circulating around a sunspot is observed.  In the H$\alpha$ blue wing it appears to be  clock-wise,  and red wing, counter-clock-wise.

\item In LB2 it shows a prominent signature of oscillations. The maximum power is at the period around 4 minutes. In LB1, the dominant motion is similar to that in LB2. However, the power is reduced by a factor of 3. In the meantime, some filament strands extend to LB1, thus the resulted component of the counter-streaming motion is due to  the physical mass motion along fibrils, resembling that in the filament.  It is possible that the oscillation is related to the surge-like oscillation above LBs as shown by Yang et al. (2015), Zhang et al. (2017) and Hou et al (2017)  using IRIS data. However, the LB under this study is much wider than those studied in the above papers.  As the observed motion is towards the solar limb (disk center) in the blue-wing (red-wing) images, the oscillation has to be perpendicular to the solar surface.

\item The magnetic field circulates around the sunspot in the clock-wise direction, connecting the LB and the filament.

\item The flux rope identified in the NLFF field shows consistency in the southern footpoint and the spinal orientation compared to the H$\alpha$ filament. The northern footpoint is slightly mis-matched.

\end{enumerate}

Based on the above observational results,  the observed counter-streaming motion in LB1 and LB2 is likely due to oscillation perpendicular to the solar surface.  The mass flows in the filament is likely due to pressure imbalance as discussed in previous literatures (e.g., Zou et al., 2016).   It is likely that the horizontal fields of part of the filament extend to LB1, evidenced by an additional component of counter-streaming motions of longer fibrils in LB1.   Consequently, some branches of the northern end of the filament are anchored in the LB1.  This mixture of oscillation and physical mass flows reduce the oscillation power in LB1.  Our study could not identify  one-to-one relationship between footpoint disturbances and the associated mass flows in episodes of counter-streaming flows in the filament.     The apparent circular counter-streaming motion is thus due to the combination of all the above effects.

\acknowledgments

We are grateful to the referee for the very helpful comments to improve the paper.  We acknowledge the BBSO and SDO teams for providing the data. This work
is supported by NSF under grants AGS 1348513, 1408703, and 1539791. R. Liu acknowledges NSFC 41474151 and 41774150 grants.

\clearpage

\begin{figure}
\epsscale{1}
\plotone{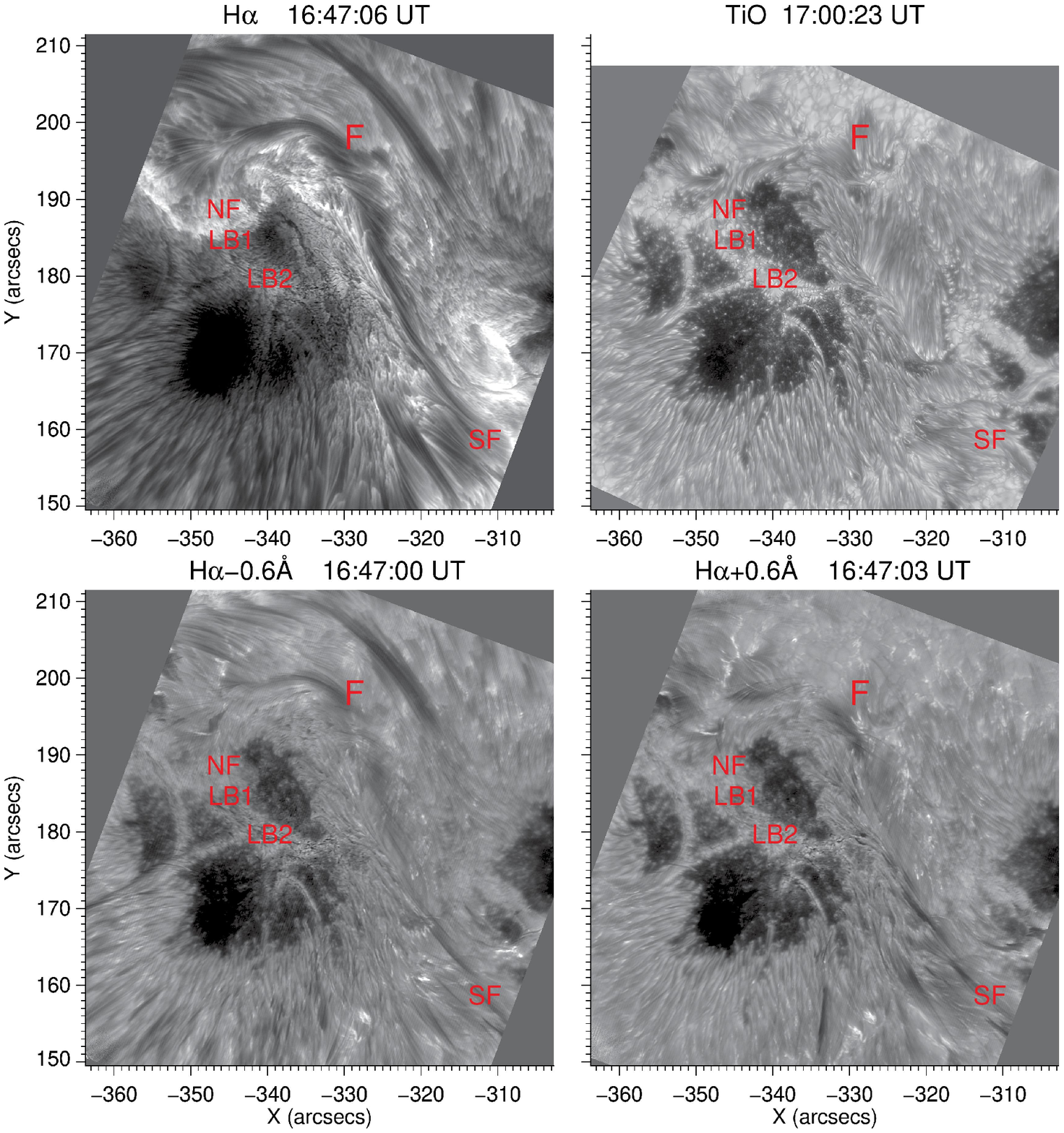}
\caption{Top-left: GST H$\alpha$ line center image, showing the main filament (F) and the locations of its northern footpoint (NF) and southern footpoint (SF). Top-right: GST TiO image, in which three sections, i.e., LB1,  LB2, and F (filament), are marked. Bottom: Corresponding H$\alpha$ $-$0.6~\AA\ and +0.6\AA\ images. Accompanied animations of blue- and red-wing time sequence images show counter-streaming motions.  \label{f1}}
\end{figure}

\begin{figure}
\epsscale{1}
\plotone{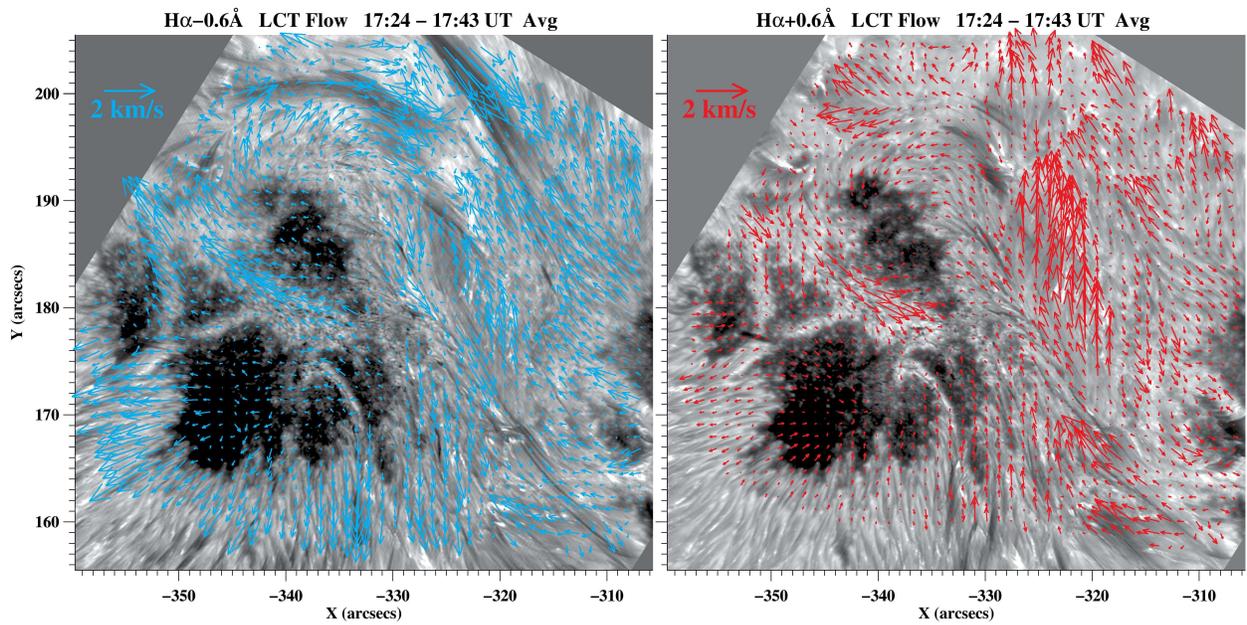}
\caption{Flows derived using the LCT method. Left: blue-wing flow map.  Right: red-wing flow map. The counter-streaming motion is apparent along F-LB1-LB2 as marked in Figure~1.  The magnitude of the flow speed is averaged over static and moving components.  \label{f2}}
\end{figure}

\begin{figure}
\epsscale{1.0}
\plotone{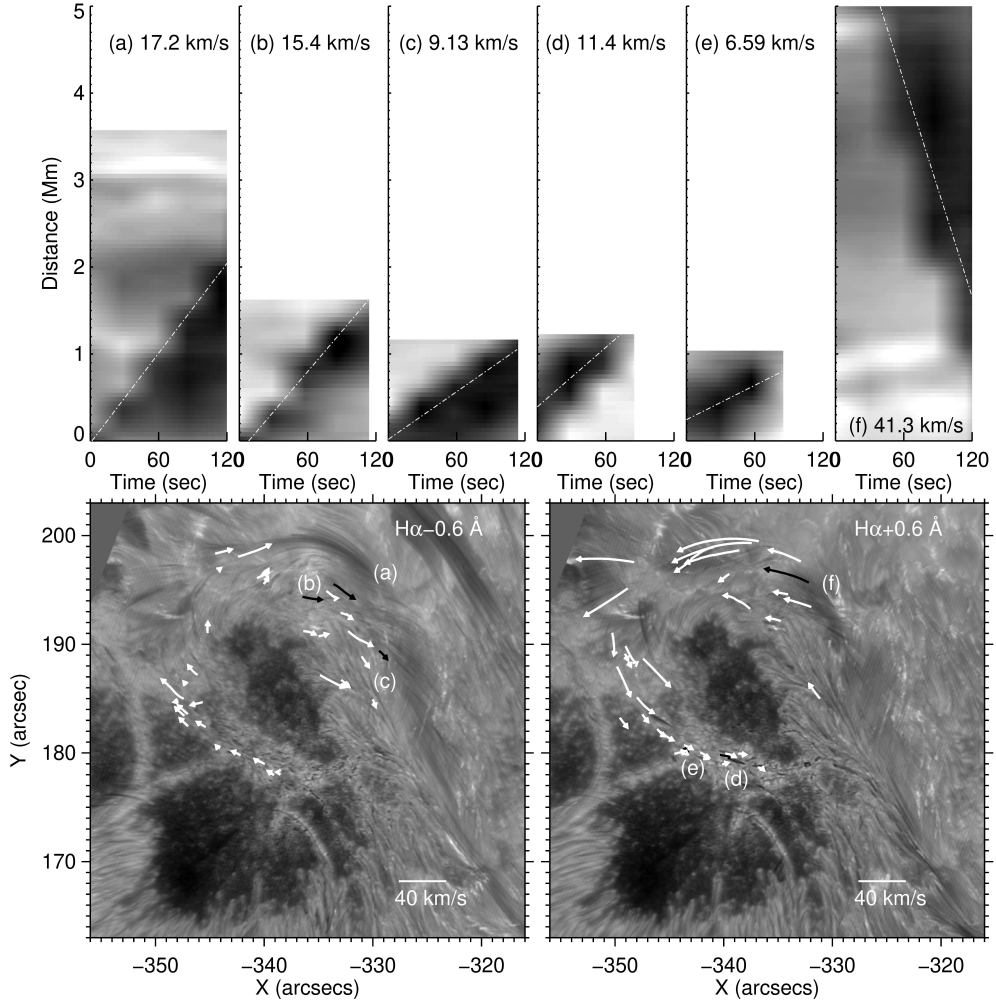}
\caption{Top: Selected examples of time-space diagrams from which the flow speeds are derived. Bottom: H$\alpha$ blue- and red-wing images at off-band of 0.6~\AA\ from the H$\alpha$ line center, over-plotted  by  flow vectors along selected H$\alpha$ fibrils.   The flow is derived based on time-distance plots. Same as Figure 2,  counter-streaming is apparent around the sunspot, connecting F, LB1 and LB2, but the flow speed is much higher. \label{f3}}
\end{figure}

\begin{figure}
\epsscale{1}
\plotone{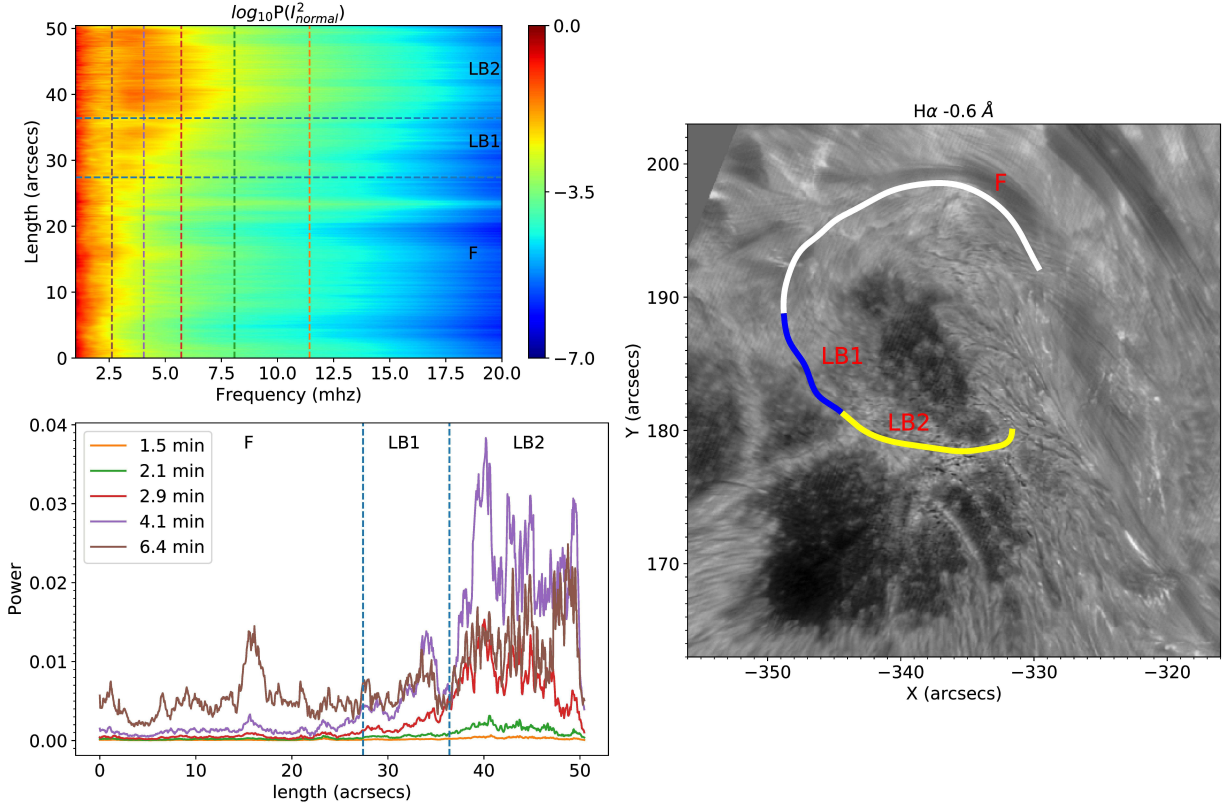}
\caption{Top-left: Oscillation power along a trajectory (defined in the right panel) that connects F, LB1 and LB2. The power is derived using wavelet analysis of pseudo Dopplergrams based on the difference between H$\alpha$ $-$0.6~\AA\ and H$\alpha$+0.6~\AA\ images.  The oscillation power in LB2 is 3 times larger than that in LB1, and more than an order of magnitude larger than the filament. This is demonstrated more clearly in the lower-left panel,  where the powers of several frequencies are plotted. The purple line is the oscillation power at a period of 4.1 minutes, at which the oscillation power peaks. \label{f4}}
\end{figure}

\begin{figure}
\epsscale{1.0}
\plotone{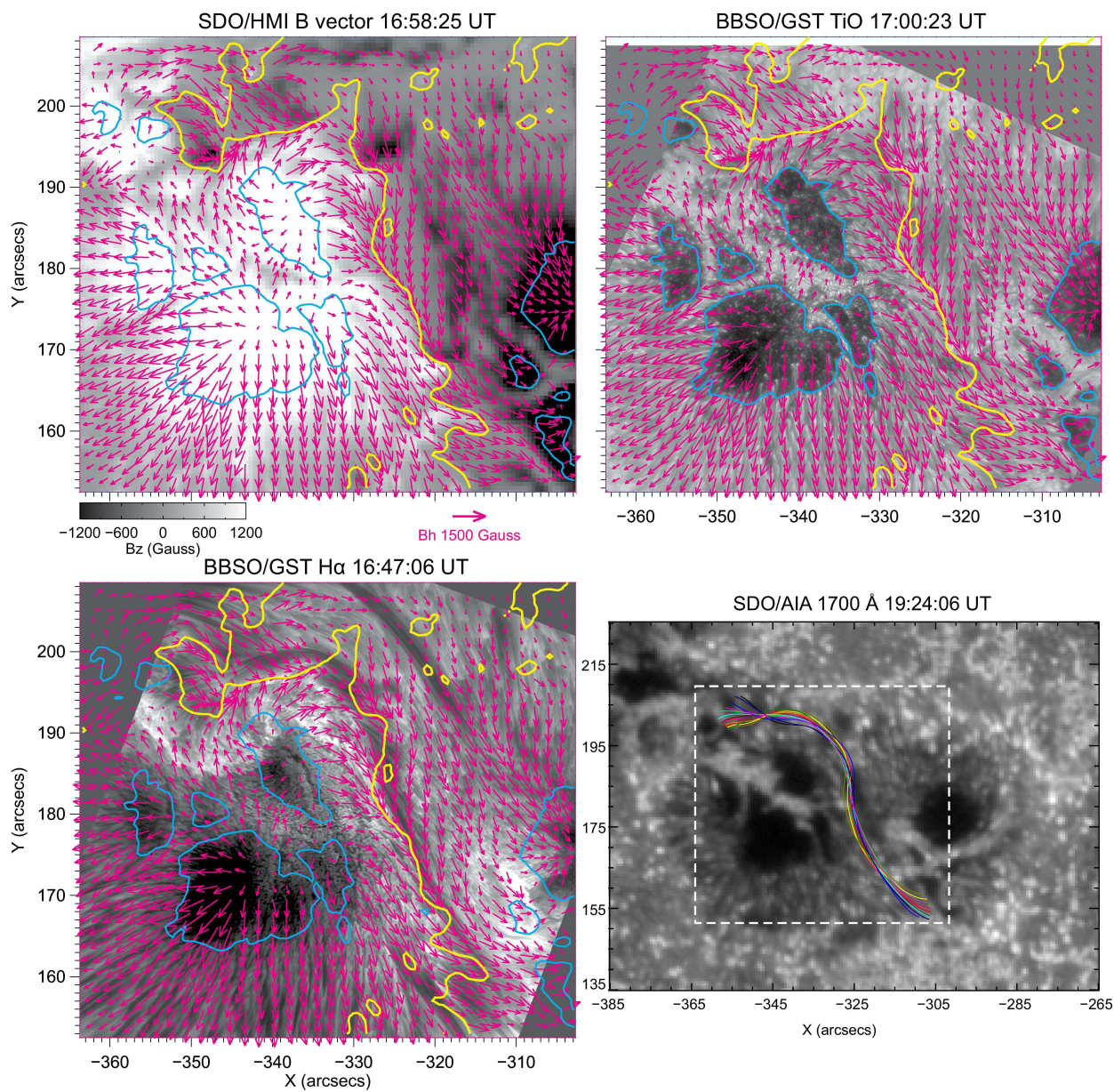}
\caption{Top-left: HMI vector magnetic field plot. Top-right and bottom-left: Transverse field over-plotted on TiO and H$\alpha$ line center images, showing the clock-wise circulation of field lines extending from the filament to LB. The blue contours mark the borders of umbrae,  while yellow contours mark the PIL. Bottom-right: Flux rope derived from NLFFF extrapolation, over-plotted on an AIA 1700~\AA\ image. The dashed box shows the FOV of the other three panels.   \label{f4}}
\end{figure}

\end{document}